\documentclass[12pt]{article}
\usepackage{graphicx}
\usepackage{hyperref}
\usepackage{amsmath}
\usepackage{amscd,amssymb}

\def\cR{{\mathcal R}}

\newcommand{\beq}{\begin{eqnarray}}
\newcommand{\eeq}{\end{eqnarray}}
\numberwithin{equation}{section}

\begin{document}


\begin{center}
{\large\bf Topological Vertex, String Amplitudes and Spectral Functions of Hyperbolic Geometry}
\end{center}

\vspace{0.1in}

\begin{center}

{\large M. E. X. Guimar\~aes $^{(a)}$ \footnote{emilia@if.uff.br},
R. M. Luna $^{{b}}$ \footnote{luna@uel.br} and T. O. Rosa $^{(a)}$
\footnote{t18rosa@if.uff.br}}

\vspace{5mm}$^{(a)}$ {\it Instituto de F\'{\i}sica,
Universidade Federal Fluminense  Av. Gal. Milton Tavares de
Souza,  s/n CEP 24210-346, Niter\'oi-RJ, Brazil}

\vspace{0.2cm} $^{(b)}$ {\it Departamento de F\'{\i}sica,
Universidade Estadual de Londrina, Caixa Postal 6001,
Londrina-Paran\'a, Brazil}

\end{center}

\vspace{0.1in}

\begin{abstract}
We discuss the homological aspects of the connection between quantum string generating function
and the formal power series associated to the dimensions of chains and homologies of
suitable Lie algebras.
Our analysis can be considered as a new straightforward application of the machinery of modular forms and spectral functions (with values in the congruence subgroup of $SL(2,{\mathbb Z})$) to the partition functions of Lagrangian branes, refined vertex and open string partition functions, represented by means of formal power series that encode Lie algebra properties.
The common feature in our examples lies in the modular properties of the characters of certain representations of the pertinent affine Lie algebras and in the role of Selberg-type spectral functions of an hyperbolic three-geometry associated with $q$-series in the computation of the string amplitudes.
\end{abstract}
\vfill



\newpage

\section{Introduction}
\label{Introduction}
In this work, we deal with some applications of the modular forms (and spectral functions related to the congruence subgroup of $SL(2, {\mathbb Z})$) to topological vertex and string generating functions. For mathematicians topological vertices  (and their respective string generating functions) may be associated to new mathematical invariants for spaces while, for physicists, they are related to quantum string partition functions.

Having given the general glance let us now explain the connection between the specific contents of the various sections in more detail. In Sect. \ref{Graded} we will explore a remarkable connection between Poincar\'{e} polynomials (generating functions)
and formal power series associated with dimensions of chains and homologies of Lie
algebras (Euler-Poincar\'{e} formula). From a concrete point of view this paper consists of applications of spectral functions to quantum string partition functions connected to suitable Lie algebras.

We show that the Poincar\'{e} polynomials (Sect. \ref{Polygraded}) and the associated topological vertex and string amplitude can be converted into product expressions which inherit cohomological properties (in the sense of characteristic classes of foliations \cite{Bytsenko3}) of appropriate  polygraded Lie algebras.

The final result for a single Lagrangian brane,
stack of branes (Sect. \ref{Single}), refined vertex, the partition functions for the case of Calabi-Yau threefolds ${\mathcal O}(-1)\oplus {\mathcal O}(-1)\mapsto \mathbb{P}^{1}$ and ${\mathcal O}(0)\oplus {\mathcal O}(-2)\mapsto \mathbb{P}^{1}$ (Sect. \ref{Vertex}) is written in terms of spectral functions of the hyperbolic three-geometry associated with $q$-series.

\section{Graded algebras and spectral functions of hyperbolic geometry}
\label{Graded}

Before considering topological vertex and string amplitudes
we would like to spend some time on relation between formal power $q$-series and homologies
of Lie algebras. We would like to show how combinatorial identities
could be derived from initial complex of (graded) Lie algebras. Our interest is the Euler-Poincar\'{e} formula associated with a complex consisting of linear spaces. The relations between Lie algebras and combinatorial identities was first discovered by Macdonald; the Euler-Poincar\'{e} formula is useful for combinatorial identities known as
Macdonald identities. The Macdonald identities are related to Lie algebras in one way or
another and can be associated with generating functions in quantum theory.

Let ${\mathfrak g}$ be an Lie algebra, and assume that it has a grading, i.e. ${\mathfrak g}$ is a direct sum of homogeneous components
${\mathfrak g}_{(\lambda)}$, where the $\lambda$'s are elements of an abelian group, $[{\mathfrak g}_{(\lambda)}, {\mathfrak g}_{(\mu)}]
\subset {\mathfrak g}_{(\lambda + \mu)}$. Let us consider a module $\bf k$ over ${\mathfrak g}$, or ${\mathfrak g}$-module, which is a left module over the universal enveloping algebra $U({\mathfrak g})$ of ${\mathfrak g}$.
Let $C^n({\mathfrak g}; {\bf k})$ be the space of $n$-dimensional cochain of the algebra ${\mathfrak g}$, with coefficients in $\bf k$.
$d= d_n : C^n({\mathfrak g}; {\bf k})\rightarrow
C^{n+1}({\mathfrak g}; {\bf k})$ as $d_{n+1}\circ d_n =0$, for all $n$, the set
$C^{\sharp}({\mathfrak g}; {\bf k}) \equiv \{C^n({\mathfrak g}; {\bf k}), d_n\}$ is an algebraic complex, while the corresponding cohomology
$H^n({\mathfrak g}; {\bf k})$ is referred to as the cohomology of the
algebra ${\mathfrak g}$ with coefficients in $\bf k$.

Let $C_n({\mathfrak g}; {\bf k})$ be the space of $n$-dimensional chains of the Lie algebra ${\mathfrak g}$ and $\delta= \delta_n:
C_n({\mathfrak g}; {\bf k})\rightarrow C_{n-1}({\mathfrak g}; {\bf k})$.
The homology $H_n({\mathfrak g}; {\bf k})$ of the complex $\{C_n({\mathfrak g}; {\bf k}), \delta_n\}$ is referred to as the homology of the algebra ${\mathfrak g}$.
We get $d(C^n_{(\lambda)}({\mathfrak g}; {\bf k})) \subset
C^{n+1}_{(\lambda)}({\mathfrak g}; {\bf k})$ and
$\delta (C_n^{(\lambda)}({\mathfrak g}; {\bf k})) \subset
C_{n-1}^{(\lambda)}({\mathfrak g}; {\bf k})$ and both spaces acquire gradings.
The chain complex
$C_{\sharp}({\mathfrak g})$,\,
${\mathfrak g}= \oplus_{\lambda=1}^{\infty}{\mathfrak g}_{(\lambda)},\,
{\rm dim}\,{\mathfrak g}_{(\lambda)}< \infty$,
can be decomposed as
\begin{equation}
0\longleftarrow C_{0}^{(\lambda)}({\mathfrak g}) \longleftarrow
C_{1}^{(\lambda)}({\mathfrak g}) \cdots\longleftarrow C_N^{(\lambda)}({\mathfrak g}) \longleftarrow 0
\end{equation}
The well known Euler-Poincar\'{e} formula reads
\begin{equation}
\sum_m(-1)^m {\rm dim}\,C_m^{(\lambda)}({\mathfrak g}) =
\sum_m(-1)^m {\rm dim}\,H_m^{(\lambda)}({\mathfrak g})\,.
\label{E-P}
\end{equation}
As a consequence, we can introduce the $q$ variable and rewrite the identity (\ref{E-P}) as a formal power
series \cite{Fuks,Bonora}:
\begin{equation}
\sum_{m,\lambda}(-1)^m q^\lambda {\rm dim}\,C_m^{(\lambda)}({\mathfrak g}) = \sum_{m,\lambda}(-1)^m q^\lambda {\rm dim}\,H_m^{(\lambda)}({\mathfrak g})
=\prod_n(1-q^n)^{{\rm dim}\,{\mathfrak g}_{n}}\,.
\label{EP}
\end{equation}

{\bf Spectral generating functions for string amplitudes.}
Having given the general scheme let us now produce, for the benefit of the reader, the
specific correspondence of the Poincar\'{e} series (which can be associated with conformal structure in two dimensionas \cite{Bytsenko2}) and spectral functions of the three-dimensional hyperbolic geometry. Thus, the infinite product identities are \cite{Bytsenko1,Bonora,Bytsenko3}:
\begin{eqnarray}
\prod_{n=\ell}^{\infty}(1- q^{\mu n+\varepsilon})
& = & \prod_{p=0, 1}Z_{\Gamma}(\underbrace{(\mu\ell+\varepsilon)(1-i\varrho(\tau))
+ 1 -\mu}_s + \mu(1 + i\varrho(\tau)p)^{(-1)^p}
\nonumber \\
& = &
\cR(s = (\mu\ell + \varepsilon)(1-i\varrho(\tau)) + 1-\mu),
\label{R1}
\\
\prod_{n=\ell}^{\infty}(1+ q^{\mu n+\varepsilon})
& = &
\prod_{p=0, 1}Z_{\Gamma}(\underbrace{(\mu\ell+\varepsilon)(1-i\varrho(\tau)) + 1-\mu +
i/(2\,{\rm Im}\,\tau)}_s
+ \mu(1+ i\varrho(\tau)p))^{(-1)^p}
\nonumber \\
& = &
\cR(s = (\mu\ell + \varepsilon)(1-i\varrho(\tau)) + 1-\mu + i/(2\,{\rm Im}\,\tau))\,,
\label{R2}
\end{eqnarray}
where $q=\exp (2\pi i\tau),\, \varrho(\tau)= {\rm Re}\tau/{\rm Im}\tau,\, \mu\in {\mathbb R}$, $\ell \in {\mathbb Z}_+$  and $\varepsilon \in {\mathbb C}$.
The Ruelle function ${\mathcal R}(s)$ is an alternating product of more complicate factors,
each of which is a Selberg-type spectral function $Z_{\Gamma}(s)$ (the analytic and modular properties of the Patterson-Selberg spectral function $Z_{\Gamma}(s)$ the reader can find in \cite{Bytsenko3}).

At this point one can use the Ruelle functions $\cR(s)$ to naturally generalize the result (\ref{R1}), (\ref{R2}) for more general infinite product identities
\begin{eqnarray}
\prod_{n=m}^{\infty}\,\big(1-q^{\mu\,n+ \varepsilon}\big)^{\nu\,n} & = &
\cR\big(s=(\mu\, m + \varepsilon)\, (1-i\varrho(\tau))+1-\mu\big)^{\nu\, m}
\nonumber \\
&\times& \prod_{n=m+1}^{\infty} \,
\cR\big(s=(\mu\, n + \varepsilon)\, (1-i\varrho(\tau))+1-\mu
\big)^{\nu}\ ,
\label{R3}
\\[4pt]
\prod_{n=m}^{\infty}\, \big(1+q^{\mu\,n+ \varepsilon}\big)^{\nu\, n} & = &
\cR\big(s = (\mu \,m + \varepsilon)\,
(1-i\varrho(\tau)) + 1-\mu+ i/(2\,{\rm Im}\,\tau)\big)^{\nu\,m}
\nonumber \\
&\times& \prod_{n=m+1}^{\infty} \,
\cR\big(s = (\mu \,n + \varepsilon)\,
(1-i\varrho(\tau)) + 1-\mu+ i/(2\,{\rm Im}\,\tau)\big)^{\nu}.
\label{R4}
\end{eqnarray}

\section{Polygraded algebras and polynomial invariants}
\label{Polygraded}

Let ${\mathfrak g}$ be a polygraded Lie algebra,
$
{\mathfrak g} = \bigoplus_{\scriptstyle \lambda_1\geq 0, ..., \lambda_k\geq 0
\atop
\scriptstyle \lambda_1+...+\lambda_k>0}
{\mathfrak g}_{(\lambda_1, ..., \lambda_k)},
$
satisfying the condition
$
{\rm dim}\, {\mathfrak g}_{(\lambda_1, ..., \lambda_k)} < \infty.
$
For formal power series in $q_1,...,q_k$, we have the following identity:
\begin{equation}
\sum_{m,\lambda_1,...,\lambda_k}(-1)^m
q_1^{\lambda_1} \cdots q_k^{\lambda_k} H_m^{(\lambda_1,...,\lambda_k)}
=
\prod_{n_1,...,n_k}\left(1-q_1^{n_1}\cdots q_k^{n_k}\right)^
{{\rm dim}\, {\mathfrak g}_{n_1, ..., n_k}}\,.
\label{poly}
\end{equation}
We would like to stress that partition functions can indeed be converted into product expressions. The expression on the right-hand side of (\ref{poly}) looks like {\it counting}
the states in the Hilbert space of a second quantized theory.

For more examples let us proceed to describing the properties of link homologies suggested by the their relation to Hilbert spaces of BPS states \cite{Gukov}.
Let ${\mathcal H}_{k,j}^{{\mathfrak{sl}(N)}; R_1, \ldots, R_\ell}(L)$
be the doubly-graded homology theory whose graded Euler characteristic
is the polynomial invariant $\overline P_{{\mathfrak sl}(N);R_1, \ldots, R_{\ell}} (q)$
(the {\it bar} means that this invariant is unnormalized invariant; its normalized version obtained by dividing by the invariant of the unknot)
\begin{equation}
\overline P_{{\mathfrak sl}(N);R_1, \ldots, R_{\ell}} (q) = \sum_{k,j \in \mathbb{Z}} (-1)^j q^k
\dim {\mathcal H}_{k,j}^{{\mathfrak{sl}(N)}; R_1, \ldots, R_\ell}(L).
\label{pslncategor}
\end{equation}
Here $L$ is an oriented link in $S^3$, we consider the Lie algebra ${\mathfrak g} = {\mathfrak sl}(N)$ (there is a natural generalization to other classical Lie algebras B, C, and D \cite{Gukov})
and a link colored is given by a collection of representations $R_1, \ldots , R_\ell$ of
${\mathfrak sl}(N)$.
In order to be agree with the standard notations we use the variables $\{q,t\}$, \cite{Dunfield}, when we discuss link homologies.

The graded Poincar\'e polynomial has the form \cite{Gukov}
\begin{equation}
\overline{{P}}_{{\mathfrak sl}(N);R_1, \ldots, R_{\ell}} (q,t) :=
\sum_{k,j \in \mathbb{Z}} q^k t^j
\dim {\mathcal H}_{k,j}^{{\mathfrak{sl}(N)}; R_1, \ldots, R_\ell}(L)\,.
\label{superpsln}
\end{equation}
By definition, it is a polynomial in $q^{\pm 1}$ and $t^{\pm 1}$
with integer non-negative coefficients. In addition,
evaluating (\ref{superpsln}) at $t=-1$ gives (\ref{pslncategor}).
In the case $R_a = \Box$ for all $a=1, \ldots, \ell$,
the homology ${\mathcal H}_{k,j}^{{\mathfrak{sl}(N)}; R_1, \ldots, R_\ell}(L)$
is known as the Khovanov-Rozansky homology, ${{}_{(KR)}\overline{H}}_{k,j}^{N} (L)$, and
its graded Poincar\'e polynomial takes the form
\begin{equation}
\overline{\mathcal{P}}_{{\mathfrak sl}(N);\Box, \cdots, \Box} (q,t)
= \sum_{k,j \in \mathbb{Z}} q^k t^j \dim {}_{(KR)}{\overline{H}}_{k,j}^N (L)\,.
\end{equation}
In order to get this expression in its final form the homology ${}_{(KR)}{\overline{H}}_{k,j}^N (L)$
has to be computed. The further physical interpretation of homological link invariants via
Hilbert spaces of BPS states leads to certain predictions regarding
the behavior of link homologies with rank $N$ (for more discussion see \cite{Gukov,Gukov1}).

\section{A single Lagrangian brane}
\label{Single}

Let us recall the connection between topological vertex and open string amplitudes in
the presence of stack of branes. It is known that in the case of a stack Lagrangian D-branes is ending on
one of the legs of the $\mathbb{C}^{3}$ (in this section and later on we will mostly follow \cite{Iqbal}
in the notations and reproduction of necessary results). The partition function is given by
\begin{equation}
{\mathcal F}(q;V)=\sum_{\nu}C_{\emptyset\,\emptyset\,\nu}(q^{-1})\,\mbox{Tr}_{\nu}V\,.
\end{equation}
Here $\mbox{Tr}_{\nu}V=s_{\nu}({\bf x})$ are the Schur functions, ${\bf
x}=\{x_{1},x_{2},\cdots\}$ are the eigenvalues of the holonomy
matrix $V$, and $C_{\lambda\mu\nu}(q)$ is the topological vertex \cite{Aganagic}.
It is well known that the topological vertex has a combinatorial interpretation in terms of counting
certain 3D partitions with fixed asymptotes \cite{Okounkov}.
Recall that Schur functions have the property that
$s_{\nu/\lambda}(Q)= Q^{|\nu|-|\lambda|}$, $\nu \succ \lambda $, and
$s_{\nu/\lambda}(Q)= 0$, otherwise.

For a single Lagrangian brane ${\bf x}=(-Q,0,0,0,\cdots)$
we get the well known partition function
\begin{equation}
{\mathcal F}(q;Q)  =  \prod_{n=1}^{\infty}(1-Q\,q^{-n+\frac{1}{2}})
\stackrel{{\rm by \,\, Eq. (\ref{R1})}}{=\!=\!=\!=\!=\!=} {\mathcal R}(s=(\alpha-1/2)(1-i\varrho(\tau))+2)\,.
\label{Ruelle}
\end{equation}
In order to be careful and to agree with the standard notations, we use the variables $\{Q, q, t\}$ when we talk about topological string amplitudes computed by the topological vertex, cf. \cite{Iqbal}. This set of variables are related with our notations $\{Q^\alpha, q= \exp(2\pi i\tau), t=\exp(2\pi i\sigma)\}$ as follows: $\tau = F_1/2\pi i,\, \sigma = F_2/F_1$, where we turn on a constant not self-dual graviphoton field strength $F = F_1dx^1\wedge dx^2+F_2dx^3\wedge dx^4$, and $\alpha = -\int_C\omega/F_1$ \cite{Iqbal}.

The above partition function of a single Lagrangian brane can
be interpreted in terms of the Hilbert series of the the symmetric
product of $\mathbb{C}$. Indeed, $s_{\nu}(Q)$ is non-zero only for those
partitions for which $\ell(\nu)=1$.
These are exactly the partitions which label the fixed points of the
symmetric product of $\mathbb{C}$, \textit{i.e.}, $\mbox{Sym}^{\bullet}(\mathbb{C})$
has a single fixed point labelled by the partition
$\nu=(\bullet,0,0,\cdots)$. A generating function of the
Hilbert series of the symmetric product is \cite{Martelli}
\begin{equation}
{\mathfrak G}(\psi,q):=\sum_{k=0}^{\infty}\psi^{k}\,H[\,\mbox{Sym}^{k}(\mathbb{C})](q).
\end{equation}
In order to determine $H[\,\mbox{Sym}^{k}(\mathbb{C})](q)$ note that the $R_k$ is the ring of symmetric functions in the variables $\{z_1, z_2, \cdots, z_k\}$ and therefore the Schur functions provide a basis of $R_k$, $R_k = \langle s_\nu(z_1,\cdots, z_k) \arrowvert \ell(\nu)\leq k\rangle$. The condition $\ell(\nu)\leq k$ is necessary since $s_\nu(z_1,\cdots, z_k)=0$ for $\ell(\nu)> k$. $R_{k}$ is isomorphic to the Hilbert space ${\mathcal H}_{k}$, generated by bosonic oscillator up to charge $k$. The Hilbert spaces $\{{\mathcal H}_{k}\}_{k=0}^\infty$ form a nested sequence
$
{\mathcal H}_{0}\subset {\mathcal H}_{1}\subset {\mathcal H}_{2}\subset {\mathcal
H}_{3}\subset \cdots
$
which corresponds to the nested sequence of Young diagrams of increasing number of rows.

The $\mathbb{C}^{\times}$ action (on $\mathbb C$ $q$ acts as a $\mathbb{C}^{\times}$ action $z\mapsto qz$), which lifts to an action on the $\mbox{Sym}^{\bullet}(\mathbb{C})$ such that the Schur functions $s_{\nu}(z_{1},\cdots,z_{k})$ are  eigenfunctions with eigenvalue
$q^{|\nu|}$,  becomes the action of $q^{L_{0}}$ on the states in ${\mathcal H}$ $(L_{0}=\sum_{n>0}\alpha_{-n}\alpha_{n}$),
\begin{eqnarray}
H[R_{k}](q) & = & \mbox{Tr}_{{\cal H}_{k}}q^{L_{0}}=\sum_{\nu|\ell(\nu)\leq
k}\,q^{|\nu|}= \prod_{n=1}^{k}(1-q^{n})^{-1}= s_{(k)}(1,q,q^{2},\cdots)
\nonumber \\
& = & \left[\frac{{\mathcal R}(s=1-i\varrho(\tau))}{{\mathcal R}(s=(k+1)(1-i\varrho(\tau)))}\right]^{-1}\!\!.
\end{eqnarray}
Note that the Hilbert series of $R_{k}$ in this case turns out to be the generating
function of the number of partitions with at most $k$ parts. Thus
the generating functions ${\mathfrak G}(\psi,q)$ is then given by
\begin{eqnarray}
{\mathfrak G}(\psi,q)&=&\sum_{k=0}^{\infty}\psi^{k}H[R_{k}](q)=\sum_{k=0}^{\infty}\psi^{k}\mbox{Tr}_{{\mathcal H}_{k}}q^{L_{0}}
= \sum_{k=0}^{\infty}\psi^{k}s_{(k)}(1,q,q^{2},\cdots)
\nonumber \\
&=&
\sum_{k=0}^{\infty}s_{(k)}(\psi)s_{(k)}(1,q,q^{2},\cdots)=\sum_{\nu}s_{\nu}(\psi)s_{\nu}(1,q,q^{2},\cdots)
= \sum_{\nu}s_{\nu}(q^{-\rho})s_{\nu}(\psi\,q^{-\frac{1}{2}})
\nonumber \\
& = & \sum_{\nu}s_{\nu^{t}}(q^{\rho})\,s_{\nu}(-q^{-\frac{1}{2}}\psi)
= \sum_{\nu}C_{\emptyset\,\emptyset\,\nu}(q^{-1})\,\mbox{Tr}_{\nu}V
= {\mathcal F}(q; Q).
\end{eqnarray}
where $\mbox{Tr}_{\nu}V=s_{\nu}(Q)$ and $Q=q^{-\frac{1}{2}}\psi$. $\mbox{Tr}_{\nu}V=s_{\nu}({\bf x})$ where ${\bf x}=\{x_{1},x_{2},\cdots\}$ are the eigenvalues of the holonomy
matrix $V$. Thus the partition function takes the form
\begin{eqnarray}
{\mathcal F}(q;V)\!\!\!\!\!\!\!&=&\!\!\!\!\!\!\!\sum_{\nu}C_{\emptyset\,\emptyset\,\nu}(q^{-1})\,s_{\nu}({\bf
x}) = \sum_{\nu}s_{\nu^{t}}(q^{\rho})s_{\nu}({\bf
x}) = \prod_{k,j=1}^{\infty}(1+q^{-k+\frac{1}{2}}x_{j})
\nonumber \\
& \stackrel{{\rm by \,\, Eq. (\ref{R2})}}{=\!=\!=\!=\!=\!=} &
\prod_{j=1}^\infty {\mathcal R}(s= (a_j-1/2)(1-i\varrho(\tau))+2+i/(2{\rm Im}\tau))\,,
\end{eqnarray}
where $a_j\equiv {\rm log}\,x_j/{\rm log}\,q$.
If we move the brane to infinity ($Q=e^{-\int_C\omega}\mapsto
0$) the contribution of the higher modes is suppressed. On the other
hand as the brane moves towards the origin ($Q\mapsto 1$) higher
oscillator modes starts contributing with equal weight to the
partition function.
It is also follows that the topological vertex $C_{\emptyset\,\emptyset\,(k)}(q)$ has an interpretation as counting the number of states of a given energy in the Hilbert space
${\mathcal H}_{k}$. It is tempting to conjecture that the topological
vertex with all three partitions non-trivial has a similar
interpretation in terms of spectral functions.

{\bf Stack of Branes.}
Let us consider the case of multiple Lagrangian branes on the
one of legs of $\mathbb{C}^{3}$. Then ${\bf x} = \{x_1, x_2, \cdots , x_N \}$
and the partition function becomes
\begin{eqnarray}
{\mathcal F}({\bf x},q)&=&\sum_{\nu}C_{\emptyset\,\emptyset\,\nu}(q^{-1})\,s_{\nu}({\bf
x}) = \prod_{j=1}^{N}\prod_{k=1}^{\infty}(1+q^{-k+\frac{1}{2}}x_{j})
\nonumber \\
& = & \prod_{j=1}^{N}{\mathcal R}(s= (a_j-1/2)(1-i\varrho(\tau))+2+i/(2{\rm Im}\tau))\,.
\label{pf}
\end{eqnarray}
It is clesr that the partition function (\ref{pf}) is the generating function of the Hilbert series of product of symmetric products of $\mathbb{C}$.

\section{Topological vertex and string amplitudes}
\label{Vertex}

{\bf Refined vertex and open string partition function.}
In this section we will explain the (combinatorial) interpretation of
the vertex in terms of 3D partitions.
The refined vertex has an interpretation in terms of generalized open topological string
amplitudes in the presence of stacks of A-brane. This vertex can be viewed as building blocks for computation of Khovanov knot invariants that can be obtained from local toric Calabi-Yau manifolds. The stack of D-branes in the context of refined
vertex can also be related to the symmetric product of $\mathbb{C}$,
as it was possible in the context of ordinary vertex. The refined topological vertex also has combinatorial interpretation
in terms of 3D partitions \cite{Iqbal}.

Note that the open string partition function depends on the leg on which the stack of branes is put. Essentially we have three choices for the open string partition function corresponding to the three legs of $\mathbb{C}^{3}$ \cite{Iqbal}:
$
{\bf (1)}\,\, C_{\lambda\,\emptyset\,\emptyset}(t,q) = (q/t)^{|\lambda|/2}\,s_{\lambda^{t}}(t^{-\rho})\,; \,\, {\bf (2)}\,\, C_{\emptyset\,\mu\,\emptyset}(t,q) = (q/t)^{(||\mu^{t}||^{2}-|\mu|)/2}\,s_{\mu^{t}}(q^{-\rho})\,; \,\,
{\bf (3)}\,\, C_{\emptyset\,\emptyset\,\nu}(t,q)=
q^{||\nu||^{2}/2}/\prod_{s\in\nu}(1-t^{1+a(s)}\,q^{\ell(s)})\,.
$

In first two cases the partition function is the same as the partition function obtained from the ordinary vertex except that the partition function depends on either $t$ or $q$
depending on the leg on which the brane ends. The third case is more interesting, the brane can ends on the prefered leg. In the case ${\bf x}= \{-Q, 0, 0, \cdots\}$ the open string amplitude is given by
\begin{eqnarray}
{\mathcal F}(Q,t,q)\!\!\!&=&\!\!\!\sum_{\nu}C_{\emptyset\,\emptyset\,\nu}(t^{-1},q^{-1})\,s_{\nu}(-Q)
=\sum_{k=0}^{\infty}C_{\emptyset\,\emptyset\,(k)}(t^{-1},q^{-1})(-Q)^{k}
\nonumber \\
& = &\!\!\!
\sum_{k=0}^{\infty}\Big(Q\frac{t}{\sqrt{k}}\Big)^{k}\,\prod_{n=1}^{k}(1-t\,q^{n-1})^{-1}
\nonumber \\
& \stackrel{t:=\, q^\sigma}{=\!=\!=\!=}& \sum_{k=0}^{\infty}\Big(Q\frac{q^\sigma}{\sqrt{k}}\Big)^{k}\left[\frac{{\mathcal R}(s= (\sigma -1)(1-i\varrho(\tau)))}{{\mathcal R}(s= (k+\sigma)(1-i\varrho(\tau)))}\right]^{-1}\,.
\end{eqnarray}
The above partition function can also be written using a more
refined Hilbert series of the symmetric product of $\mathbb{C}$.

{\bf The case of the Calabi-Yau threefold ${\mathcal X}={\mathcal O}(-1)\oplus {\mathcal O}(-1)\mapsto \mathbb{P}^{1}$.} As before one can use the refined topological vertex to
determine the generalized partition function for various local toric
Calabi-Yau threefolds. The compactification of Type IIA string theory on the
${\mathcal X}\mapsto \mathbb{P}^{1}$ gives rise to $U(1)$ ${\mathcal N}=2$ gauge theory on
the transverse $\mathbb{C}^{2}$ in a particular limit \cite{Boyer}.
The topological string partition function is given by
\begin{eqnarray}
{\mathcal F}(q,Q)\!\!\!\!\!&=&\!\!\!\!\!\sum_{\nu}Q^{|\nu|}(-1)^{|\nu|}\,C_{\emptyset\,\emptyset\,\nu}(q)\,C_{\emptyset\,\emptyset\,\nu^{t}}(q)
= \sum_{\nu}Q^{|\nu|}(-1)^{|\nu|}s_{\nu^{t}}(q^{-\rho})s_{\nu}(q^{-\rho})
\nonumber \\
&=&\!\!\!\!\!\prod_{k,j=1}^{\infty}\Big(1-Q\,q^{k+j-1}\Big)=\prod_{n=1}^{\infty}\Big(1-q^{n}\,Q\Big)^{n}
\nonumber \\
& \stackrel{{\rm by \,\, Eq. (\ref{R3})}}{=\!=\!=\!=\!=\!=} &
\prod_{n=1}^\infty {\mathcal R}(s= (n+ \alpha)(1-i\varrho(\tau)))\,.
\end{eqnarray}
where $-\mbox{log}\,Q = \int_C\omega$ is the K\"{a}hler parameter, the size
of the $\mathbb{P}^{1}$. Thus the refined topological vertex can be used to determine the refined
partition function. The other choice of the refined partition function is
\begin{eqnarray}
{\mathcal F}(t,q,Q)&=&\sum_{\lambda}Q^{|\lambda|}(-1)^{|\lambda|}\,C_{\lambda\,\emptyset\,\emptyset}(t,q)\,
C_{\lambda^{t}\,\emptyset\,\emptyset}(q,t)
=
\sum_{\lambda}(-Q)^{|\lambda|}\Big(\tfrac{q}{t}\Big)^{\frac{|\lambda|}{2}}\,
s_{\lambda^{t}}(t^{-\rho})\,\Big(\tfrac{t}{q}\Big)^{\frac{|\lambda|}{2}}\,s_{\lambda}(q^{-\rho})\,
\nonumber \\
&=&\sum_{\lambda}(-Q)^{|\lambda|}s_{\lambda^{t}}(t^{-\rho})\,s_{\lambda}(q^{-\rho})\,=\prod_{k,j=1}^{\infty}(1-Q\,q^{k-\frac{1}{2}}\,t^{j-\frac{1}{2}})\,
\nonumber \\
&=&
\prod_{j=1}^\infty{\mathcal R}(s= (\alpha + \sigma(j-1/2)-1/2)(1-i\varrho(\tau)))\,.
\end{eqnarray}

{\bf The case of the Calabi-Yau threefold ${\mathcal X}={\mathcal O}(0)\oplus {\mathcal O}(-2)\mapsto \mathbb{P}^{1}$.}
One can obtain this geometry from local $\mathbb{P}^{1}\times
\mathbb{P}^{1}$ by taking the size of one of the $\mathbb{P}^{1}$
very large  and get two copies of ${\mathcal O}(0)\oplus {\mathcal
O}(-2)\mapsto \mathbb{P}^{1}$.

In the topological vertex formalism the partition functions take the form
\begin{eqnarray}
{\mathcal F}(q,Q)&=&\sum_{\nu}Q^{|\nu|}(-1)^{|\nu|}C_{\emptyset\,\emptyset\,\nu}(q)\,(-1)^{|\nu|}\,q^{\frac{\kappa(\nu)}{2}}\,C_{\emptyset\,\emptyset\,\nu^{t}}(q)
= \sum_{\nu}Q^{|\nu|}s_{\nu^{t}}(q^{-\rho})\,q^{\frac{\kappa(\nu)}{2}}\,s_{\nu}(q^{-\rho})
\nonumber \\
& = & \sum_{\nu}Q^{|\nu|}s_{\nu^{t}}(q^{-\rho})\,s_{\nu^t}(q^{-\rho})
= \prod_{k,j=1}^{\infty}\Big(1-Q\,q^{k+j-1}\Big)^{-1}\,=\prod_{n=1}^{\infty}\Big(1-Q\,q^{n}\Big)^{-n}
\nonumber \\
&=&
\prod_{n=1}^\infty\left[{\mathcal R}(s= (n+\alpha)(1-i\varrho(\tau)))\right]^{-1}\,.
\\
{\mathcal F}(t,q,Q)&=&\sum_{\lambda}Q^{|\lambda|}(-1)^{|\lambda|}\,C_{\emptyset\,\lambda\,\emptyset}(t,q)\,f_{\lambda}(t,q)\,\,
C_{\lambda^{t}\,\emptyset\,\emptyset}(t,q)
\nonumber \\
& = &
\sum_{\lambda}(-Q)^{|\lambda|}\Big(\frac{q}{t}\Big)^{\frac{||\lambda||^2}{2}}
t^{\frac{\kappa(\lambda)}{2}}\,s_{\lambda^{t}}(q^{-\rho})\,f_{\lambda}(t,q)\,s_{\lambda}(t^{-\rho})
\nonumber \\
&=&\sum_{\lambda}(Q\sqrt{\tfrac{q}{t}})^{|\lambda|}\,s_{\lambda^{t}}(t^{-\rho})\,s_{\lambda^{t}}(q^{-\rho})
=\prod_{k,j=1}^{\infty}\Big(1-Q\,q^{k}\,t^{j-1}\Big)^{-1}
\nonumber \\
&=&
\prod_{j=1}^\infty\left[{\mathcal R}(s= (1+\alpha+\sigma(j-1))(1-i\varrho(\tau)))\right]^{-1}\,.
\end{eqnarray}

\section{Conclusions}

The main result and the central concept in all examples here are that the quantum generating functions considered in this paper can be converted into products of spectral functions associated with $q$-series.
This common feature encodes the connection with infinite-dimensional Lie algebras and their homologies, together with the remarkable link to hyperbolic geometry.
The Poincar\'{e} polynomials and the associated topological vertices and string amplitudes can be converted into product expressions which inherit modular and homological properties (in sense of characteristic classes of foliations, as it has been stressed for the first time in \cite{Bytsenko3}) of appropriate polygraded algebras.

\subsection*{Acknowledgements}

The authors  would like to thank Prof. A. A. Bytsenko for quite useful discussions and suggestions during the preparation of this manuscript. The authors thanks CAPES, CNPq and
Funda\c{c}\~ao Arauc\'aria (Paran\'a, Brazil) for financial support. Finally, the authors also would like to thank the referees for important criticisms which improved a lot the previous manuscript.

\end{document}